\documentstyle[12pt,epsf,aaspp4]{article}

\def\kms{{\rm\,km/s}}
\def\msun{{\rm\,M_\odot}}

\def\vol#1  {{{#1}{\rm,}\ }}

\def\etal{et al.\ }

\newcount\refno
\newcount\rfno
\def\eq{$^{\the\refno\ }$\advance\refno by 1}
\def\ad{\advance\rfno by 1}

\def\clock{\count0=\time \divide\count0 by 60
        \count1=\count0 \multiply\count1 by -60
\advance\count1 by
\time

\number\count0:\ifnum\count1<10{0\number\count1}
\else\number\count1\fi}

\def\kms{\rm km/s}
\def\gcm{{\rm g/cm}}

\def\Gcm2{\rm G~cm^2}

\def\beq{\begin{equation}}
\def\eeq{\end{equation}}

\def \date         {\ifcase\month \message{zero}
\or
                       January \or February \or
March \or April \or May \or June
                       \or July \or
                       August \or September \or
October \or November
\or
                       December \fi
                       \space\number\day,
\number\year}

\begin{document}

\title{Origin of the Black Hole Mass - Bulge Velocity Dispersion Relation: 
A Critical Surface Density For Disk Accretion}
\author{Renyue Cen\altaffilmark{1}}

\altaffiltext{1} {Princeton University
Observatory,  Princeton University, Peyton Hall, Ivy Lane, Princeton, NJ
08544; cen@astro.princeton.edu.}
\received{\date}
\accepted{ }

\begin{abstract}
It is shown that, if gas accretion via a disk onto the central supermassive black hole 
is efficient only for surface density $\Sigma \ge 10$g/cm$^2$,
the black hole mass - galactic bulge velocity dispersion relation (Tremaine \etal 2002)
is naturally borne out,
so is the modest dispersion in that relation,
in the context of hierarchical structure formation theory.
This relation is not expected to evolve with redshift in this model.

\end{abstract}

\keywords{accretion disks --- black hole physics --- galaxies:formation
galaxies - star formation}

\section{INTRODUCTION}

Both the observed correlation between the mass of the central supermassive
black hole (SMBH) and the velocity dispersion of the host galactic bulge,
and the small dispersion about that correlation (Tremaine \etal 2002),
are intriguing and not well understood.
Several interesting models have been offered to possibly provide 
an explanation for such a relation (Silk \& Rees 1998;
Ostriker 2000;
Adams, Graff, \& Richstone 2001; 
Colgate \etal 2003).
In this {\it Letter} we provide an alternative model and
make the case that,
if there is a critical surface density for accretion disks at 
$\Sigma \sim 10$g/cm$^2$,
then the exact observed relation as well as
the modest dispersion in the relation can be obtained
in the cold dark matter model.

\section{Synchronous Growth of Supermassive Black Hole and Bulge}
\label{sec:growth}

During a significant merger event
gravitational torques drive a significant amount gas 
towards the central region (Barnes \& Hernquist 1991; 
Mihos \& Hernquist 1996).
What is the density run of the resulting gas disk?

Simulations have shown that $M(<j) \propto j$ for dark matter halos
(Bullock \etal 2001; Cen \etal 2004) at the low $j$ end,  
where $M(<j)$ is the amount of matter with specific angular momentum smaller
than $j$.
It seems reasonable to expect that such a relation may be extended
to gas in halos, since both dark matter and gas
are subject to largely the same gravitational forces
(van en Bosch \etal 2002).
When the gas in the small $j$ end is channeled to the central region
during the merger and cools, 
with lower $j$ gas self-adjusting to settle at smaller radii,
it would produce a self-gravitating gas disk 
whose surface density runs as:
\begin{equation} 
\Sigma(r) = \Sigma_0 (r/r_0)^{-1}.
\end{equation}
\noindent 
This occurs when a pre-existing gravitational potential well 
is small compared to that produced by the newly formed gas disk.
On the other hand, if the newly formed gas disk only incrementally 
fortifies an existing gravitational potential well,
the gas disk should also follow Equation (1) in a steady state,
since the existing gravitational potential well corresponds to 
a flat rotation curve.
Thus, it seems that a flat rotation curve may be maintained 
in the galactic bulge
through a sequence of largely random mergers.
Indeed, observations indicate that the rotation curves
in galactic bulges are nearly flat within a factor of two over a radial span 
of $3-4$ decades (Sofue \& Rubin 2001),
except in the very inner region where the gravitational influence of the central
SMBH dominate.
These considerations lead us to conclude that 
it may be assumed, with good accuracy,
that the velocity dispersion $\sigma$ within the bulge region is constant  
and the infalling gas from each merger event 
produces a gas disk of a surface density run governed by Equation (1).
The exact value of $\Sigma_0$ at a fixed $r_0$ will depend on
the strength of the merger event.
We note that both $\Sigma_0$ and $r_0$ cancel out in the final
expression that we derive below.

Following a merger, when the disk gas within a physical radius $r_0$ forms into stars 
(except the very inner small disk which will be assumed to accrete
mostly onto the SMBH and is a very small fraction of the overall mass),
an amount of stellar mass equal to 
\begin{equation} 
\Delta M_* = 2\pi \Sigma_0 r_0^2
\end{equation}
\noindent 
will be added to the bulge within $r_0$.

Let us now propose that only the gas in the inner region where the surface density 
exceeds $\Sigma_c$ at a radius $r_c(\Sigma_0,r_0)$
will be accreted onto the central 
SMBH. It can then be shown that the following mass,
\begin{equation} 
\Delta M_{BH} = {(\Delta M_*)^2 \over 2\pi r_0^2 \Sigma_c},
\end{equation}
\noindent 
will be added to the SMBH due to gas accretion.
To make things simple, we assert that 
$\Delta M_{BH}$ in Equation (3) holds,  
even if $r_c$ is smaller than the Bondi radius,
with which SMBH may start
to dominate the gravitational potential well and the rotation velocity rises.
We assume that there, instead, will be an initial, unsteady 
state of gas accretion onto the SMBH in this case, until 
a steady state is reached that conforms to the rotation velocity profile.
In a hierarchical cold dark matter structure formation model
(Spergel \etal 2006) galaxies grow through mergers and acquisitions
(Lacey \& Cole 1993; Kauffmann, White, \& Guiderdoni 1993).
Equation (3) suggests that the central SMBH
grow synchronously with the bulges of galaxies,
albeit at a different rate: the growth of black holes
is more heavily facilitated by large mergers.
Integrating equation (3) yields 
\begin{equation} 
M_{BH} = {\sum (\Delta M_*)^2 \over 2\pi r_0^2 \Sigma_c} + C,
\end{equation}
\noindent 
where $\sum (\Delta M_*)^2$ denotes the sum of the squares
of stellar mass increments in the bulge interior to $r_0$ from mergers over the entire 
growth history.
Since the initial mass of the central black hole is zero or small, if there is 
a small seed black hole,
the integration constant $C$ is, for all practical purposes, zero.
We rewrite equation (4) as 
\begin{equation} 
M_{BH} = {g(M_*) M_*^2 \over 2\pi r_0^2 \Sigma_c},
\end{equation}
\noindent 
where we have defined 
\begin{equation} 
g(M_*) \equiv {\sum (\Delta M_*)^2 \over M_*^2}, M_* \equiv \sum \Delta M_*,
\end{equation}
\noindent 
where $M_*$ is the stellar mass within $r_0$ at the redshift under consideration.
Note that the left-hand side of Equation (5) automatically includes
contributions of pre-existing BHs in merging galaxies.
Equation (5) may be cast into the following form:
\begin{equation} 
M_{BH} = {2 g(M_*) \sigma^4 \over \pi G^2 \Sigma_c} 
= 1.3\times 10^8 \left({\sigma\over 200\kms}\right)^4\left({\Sigma_c\over 10 \gcm^2}\right)^{-1} \left({g(M_*)\over 0.11}\right) \msun,
\end{equation}
\noindent 
where $\sigma$ is 1-d velocity dispersion of the bulge and $G$ is the gravitational
constant.
Equation (7) would be in a remarkably good agreement with 
the observed one,
$M_{BH} =  (1.3\pm 0.2)\times 10^8 \left({\sigma\over 200\kms}\right)^{4.02\pm 0.32}\msun$
(Tremaine \etal 2002),
if the two parameters, $g$ and $\Sigma_c$,
in the equation have the fiducial constant values given.

All cosmological uncertainties in $M_{BH}$ have now been condensed into $g(M_*)$,
which depends on the assembly history of a bulge,
for a given $M_*$; i.e., $g(M_*)$ may not only depend on $M_*$ but also
have multiple values (i.e., a dispersion) at a fixed $M_*$.
Since $g(M_*)$ is not easily computable,
we instead will make the assumption that the amount of 
gas driven towards the central region in a merger
is proportional to the strength of a merger event times the host mass.
A precise definition of the merger strength is, however, difficult to pin down.
The mass ratio of the merger pair, $R=M_1/M_2$ (where $M_1$ and $M_2$
denote the mass of the small and large halo of the merging pair, respectively), 
impact parameter, orbital inclination,
initial orbital energy, the sizes of the halos, 
a pre-existent bulge, etc., may all play a role to a varying extent (Mihos \& Hernquist 1996).
Instead, we will simply use $R$ as a proxy to 
characterize a merger strength.
Since the neglected factors that may be involved, such as the orbital inclination,
a pre-existing bulge, may largely behave like random variables in the overall
growth history of a galaxy,
it may be a good approximation to absorb all factors into one factor,
$R$, in this case, which is known to be perhaps the most important.
As will be shown below, the exact definition of 
the merger strength in terms of $R$ do not appear to matter as far as $g$ is concerned.
We also assume that 
there is a threshold (lower bound), $R_{th}$, in order to 
drive a significant amount of gas towards the central region.
Again, it turns out that the results do not materially
depend on $R_{th}$.
We would like to point out that a merger threshold
may be operating, because there exist bulge-less spiral galaxies,
indicating that the growth of a galaxy as a whole
does not necessarily result in a bulge.

\begin{figure}
\figurenum{1}
\plotone{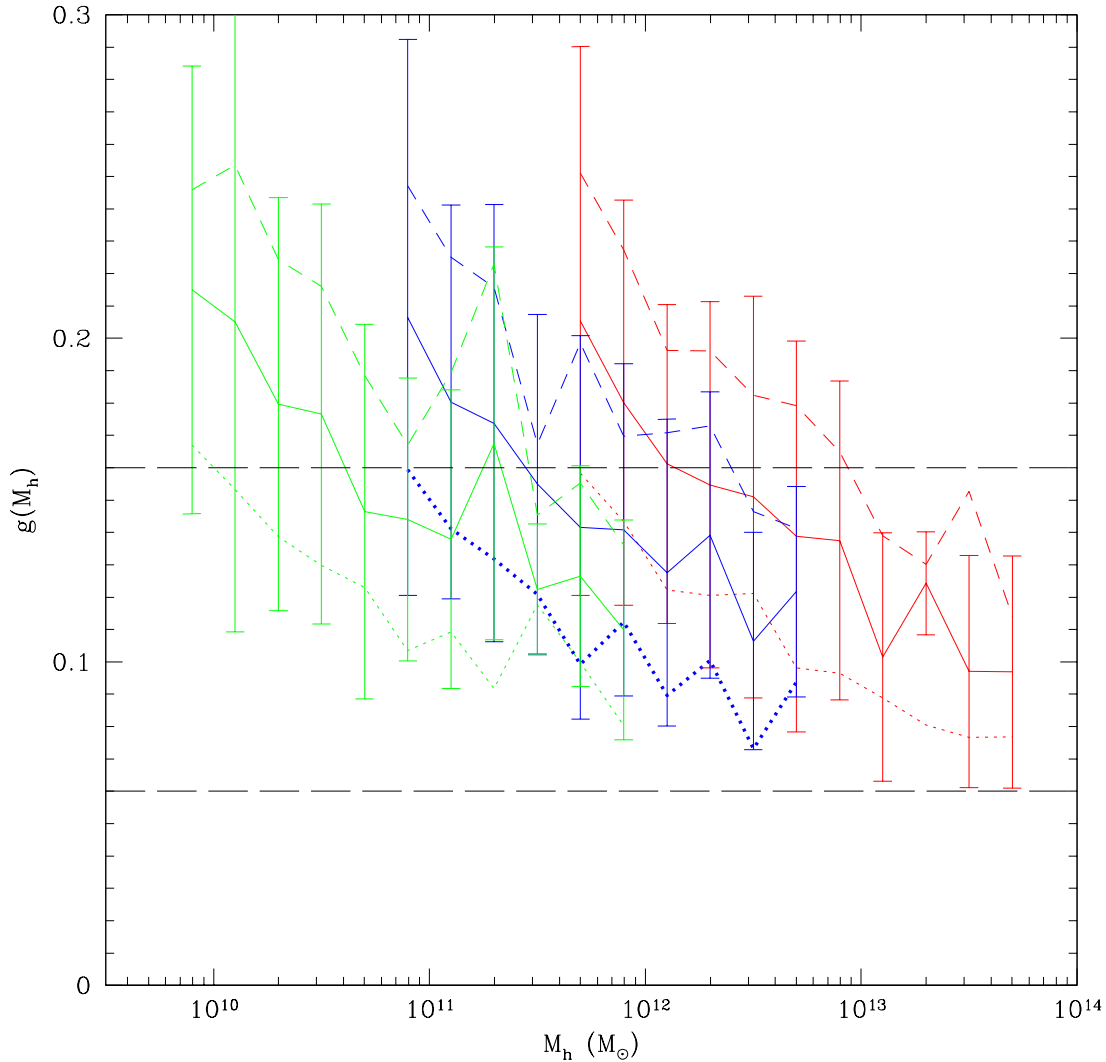}
\caption{
shows $g(M)$ as a function of halo mass $M_h$.
We assume that each merger channels an amount of gas 
$\Delta M_* \propto M_2 R$ for $R\equiv M_1/M_2\ge R_{th}$ and zero otherwise, 
where $M_1$ and $M_2$ denote the mass of 
the small and large halo of the merging pair, respectively.
Three simulations are run each with $512^3$ particles, in a box of size
$(50,25,12.5)$Mpc/h comoving, respectively, shown from right to left 
in red, blue and green.
The dotted, solid and dashed sets of curves have $R_{th}=(0.1,0.2,0.3)$, respectively.
$1\sigma$ variance is displayed only for the case with $R_{th}=0.2$.
The three simulations (and some additional ones, not shown here)
are run to test resolution effects.
It is clear that the upturn towards small mass for each curve
is resolution effect; low mass ratio merger events in small halos hence $g(M_h)$
are underestimated due to resolution effect.
The two horizontal long dashed lines
approximately bracket the range of $g=0.11\pm 0.05$.
The standard cold dark matter model with a cosmological constant
using the parameters determined by WMAP3 (Spergel \etal 2006) is used.
}
\label{f1}
\end{figure}

\begin{figure}
\figurenum{1}
\plotone{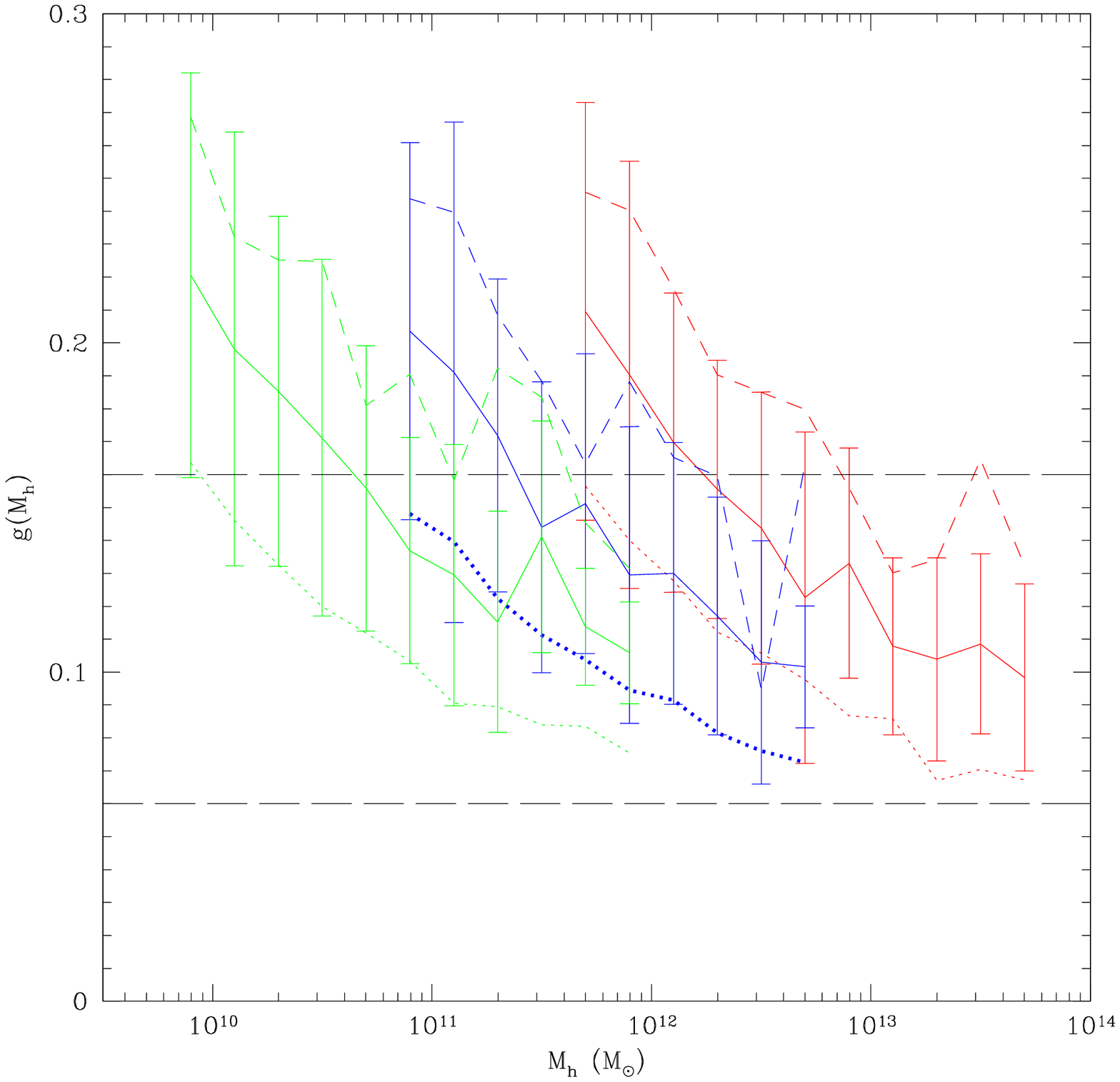}
\caption{
is similar to Figure 1, except that we 
assume each merger channels an amount of gas 
$\Delta M_* \propto M_2\times R^2$.
}
\label{f2}
\end{figure}

\begin{figure}
\figurenum{1}
\plotone{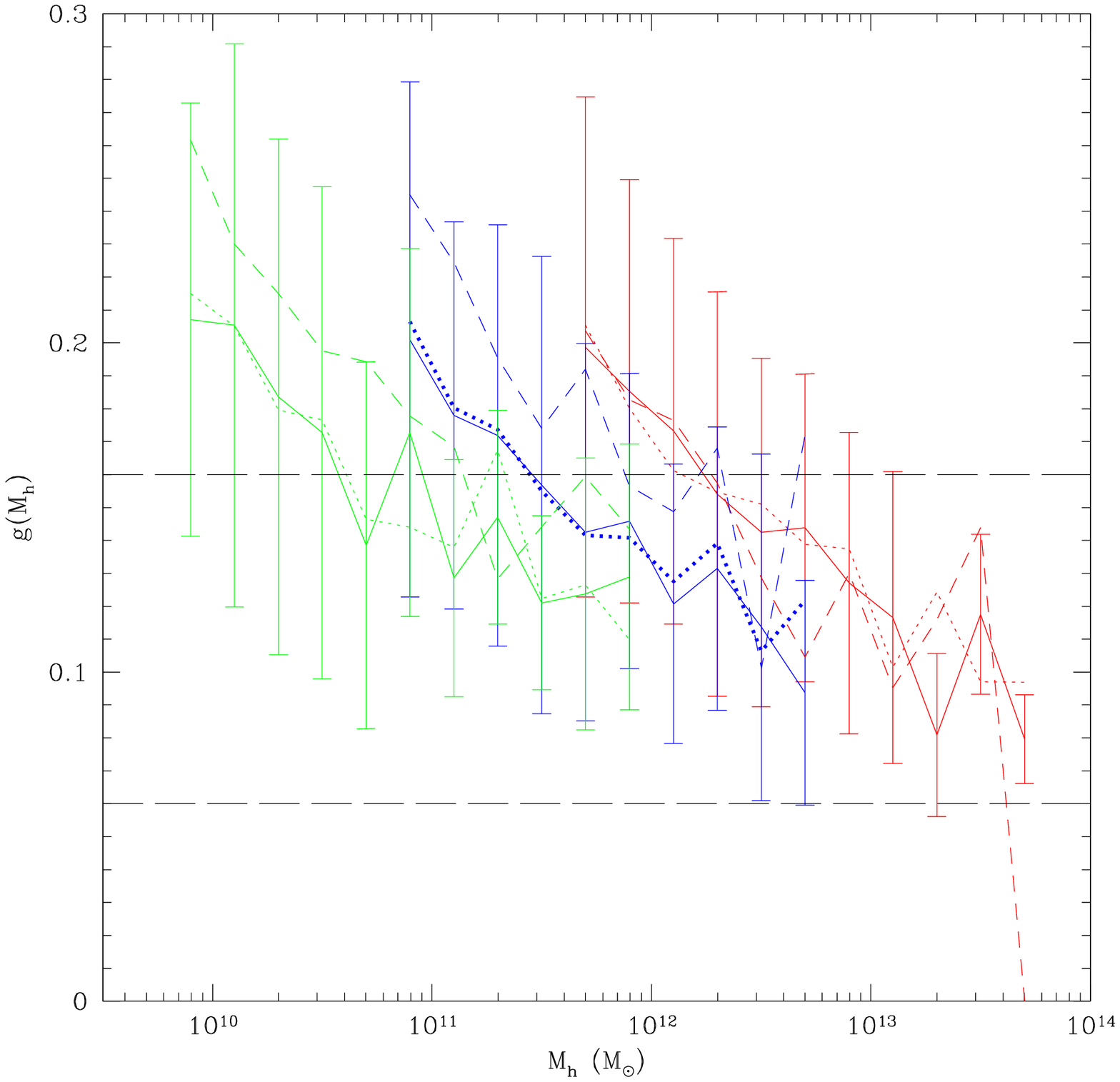}
\caption{
shows $g(M)$ as a function of halo mass $M_h$ at $z=0,1,2$ with dotted,
solid and dashed curves, respectively, all with $R_{th}=0.2$,
with the assumption that $\Delta M_* \propto M_2\times R$.
}
\label{f2}
\end{figure}

Figure 1 shows $g(M_h)$ as a function of halo mass $M_h$,
where we assume that each merger channels an amount of gas 
$\Delta M_* \propto M_2\times R$. 
In Figure 2 we use a different dependence of the amount
of gas that is driven to the center on the merger mass ratio,
$\Delta M_* \propto M_2\times R^2$.
We see that $g(M_h)$ depends weakly on $M_h$.
The dependences of $g(M_h)$ on both $R$ and $R_{th}$ are also weak. 
Note that each curve from each simulation box tends to rise
at the low halo mass end.
We understand that this is a resolution effect.
Basically, low mass ratio merger events in small halos hence $g(M_h)$
are underestimated due to resolution effect.
Since the overall stellar mass increase in the bulge 
is linearly proportional to the stellar mass increase $\Delta M_*$
while the increase in SMBH mass depends quadratically on  $\Delta M_*$,
$g$ should artificially rise if small $\Delta M_*$ are not accounted for.
Our extensive testing verified this, which is easily seen in Figure 1
by comparing the results from three shown simulation boxes.

The results are consistent with $g$ being constant,
$g=0.11\pm 0.05$ (shown as the two horizontal long dashed lines
in Figures 1,2), bracketing the computed range of halo masses 
where merger-tree simulation results are deemed to be reliable
and for $R_{th}=0.1-0.3$.
The relatively small dispersion in $g$ and its near constancy
with respect to $M_h$ may be reflective of temporal
smoothing of the growth of a black hole/bulge by 
a sequence of largely ``random" merger events of varying strengths
and a relatively featureless cold dark matter power spectrum.

Equation (7) additionally indicates that the dispersion in
the mass of SMBH at a fixed bulge velocity dispersion
may also be contributed by a dispersion in $\Sigma_c$,
which may be of astrophysical origin.
The fact that the observed dispersion in $M_{BH}-\sigma$ relation
is no larger than $\sim 0.3$ dex (Tremaine \etal 2002) 
implies that the uncertainty $\Sigma_c$ should be no larger than 
a factor of about two.
In other words, it seems that 
whatever physics dictating $\Sigma_C$ operates in a precision fashion.
We see that if $\Sigma_c\sim 10$g/cm$^2$, 
the observed $M_{BH}-\sigma$ relation (Tremaine \etal 2002)
is produced.

The derived result, Equation (7), 
appears to represent a ``fundamental" line,
relating $M_{BH}$ to a dynamical quantity of the bulge,
in this case, the velocity dispersion $\sigma$.
It might be that, if one were to express $M_{BH}$ against, say,
the total stellar mass in the bulge $M_*(tot)$, 
the scatter would be significantly larger.
This seems to be expected given the nonlinear
dependence of $\Delta M_{BH}$ on $\Delta M_*$ (Equation 3).

Figure 2 shows $g(M_h)$ for $z=(0,1,2)$.
We see that $g$ does not evolve significantly in this redshift range.
We do not show results at higher redshifts due to merger-tree 
simulation particle mass resolution effect
and significantly larger cosmic variances.
But there is no indication that $g$ will not be constant
and will be significantly different at higher redshifts.
Therefore, we predict that the local observed BH mass - bulge velocity dispersion relation 
(Tremaine \etal 2002) is expected to hold at all redshifts.
This provides a unique and critical test of the model.

\section{DISCUSSION AND CONCLUSIONS}
\label{sec:discuss}

We have proposed a simple model for the growth of SMBH based on
the hierarchical formation of galaxies and a critical assumption
that gas accretion onto SMBH is only effective for surface densities
in excess of $\Sigma \sim 10$g/cm$^2$.
We show that the exact observed relation between the mass of the SMBH and
the velocity dispersion of the bulge (Tremaine \etal 2002) 
is obtained in this model (Equation 7).
Equally important, the small dispersion in the relation
may be expected in this model.
It appears that this $M_{BH}-\sigma$ relation 
represents a ``fundamental" line,
relating $M_{BH}$ to the dynamic state of the bulge
(the velocity dispersion $\sigma$).
A definitive additional prediction is that the 
$M_{BH}-\sigma$ relation does not evolve with redshift in this model,
which provides a test of the model.

On the other hand, we show that, while the overall growth of SMBH 
is roughly synchronous with the growth of their host galactic bulges, 
it has a different rate and in general the increase in mass of the SMBH
is not necessarily proportional to the increase in bulge mass.
As a result, it is expected that the correlation between the mass of SMBH
and the overall bulge mass may display a much large scatter.

The physical origin for such a critical surface density
in the standard accretion disk (Shakura \& Sunyaev 1973) is,
however, still unclear.
A general requirement may be that gas accretion onto the central SMBH 
is dominant over star formation at $\Sigma \ge 10$g/cm$^2$ and the reverse 
is true at $\Sigma \le 10$g/cm$^2$.
It is interesting to note that $\Sigma_c\sim 10$ corresponds to
an optical depth of $\sim 10$.
Therefore, $\Sigma_c$ may be directly or indirectly related to 
a transition from an optically thin to optically thick disk.
The accretion disk model based on the large-scale vortices 
of the Rossby vortex instability may offer one solution (Colgate \etal 2003).
The possible existence of an unstable disk at some large radii, 
which is expected to be prone to star formation, 
may offer another solution (Goodman 2003).
Further investigations are warranted.

\acknowledgments
I thank Rashid Sunyaev for a stimulating conversation.
I am grateful to Pierluigi Monaco for providing a merger 
tree code (Pinocchio) and useful correspondences regarding the
usage of the code.
This work is supported in part by grants NNG05GK10G and AST-0507521.

\end{document}